\documentclass[prl
,floatfix
,preprintnumbers
,amsmath
,amssymb
,superscriptaddress]{revtex4}
\usepackage[T1]{fontenc}
\usepackage{graphicx}
\usepackage{dcolumn}
\usepackage{bm}
\usepackage[dvips]{color}

\begin{document}

\title{Effect of fingerprints orientation on skin vibrations\\ during tactile exploration of textured surfaces}

\author{A. Prevost}
\email{alexis.prevost@lps.ens.fr.}
\affiliation{Laboratoire de Physique Statistique de l'ENS, UMR 8550, CNRS-ENS-Université Paris 6 \& Paris 7, 24 rue Lhomond F-75231 Paris, France.}
\author{J. Scheibert}
\affiliation{Physics of Geological Processes, University of Oslo, Oslo, Norway}
\author{G. Debrégeas}
\affiliation{Laboratoire de Physique Statistique de l'ENS, UMR 8550, CNRS-ENS-Université Paris 6 \& Paris 7, 24 rue Lhomond F-75231 Paris, France.}

\begin{abstract}
In humans, the tactile perception of fine textures is mediated by skin vibrations induced by scanning the surface with the fingertip. These vibrations are encoded by specific mechanoreceptors, Pacinian corpuscules (PCs), located about 2mm below the skin surface. In a recent article, we performed experiments using a biomimetic sensor which suggest that fingerprints (epidermal ridges) may play an important role in shaping the subcutaneous stress vibrations in a way which facilitates their processing by the PC channel. Here we further test this hypothesis by directly recording the modulations of the fingerpad/substrate friction force induced by scanning an actual fingertip across a textured surface. When the fingerprints are oriented perpendicular to the scanning direction, the spectrum of these modulations shows a pronounced maximum around the frequency $v$/$\lambda$, where $v$ is the scanning velocity and $\lambda$ the fingerprints period. This simple biomechanical result confirms the relevance of our previous finding for human touch.\\

\noindent Addendum to: Scheibert J, Leurent S, Prevost A, Debrégeas G. The role of fingerprints in the coding of tactile information probed with a biomimetic sensor. Science 2009; 323:1503-1506.\\

\end{abstract}

\maketitle

 The skin plays a major role in the sense of touch. It carries information from the external world to the embedded mechanoreceptors \cite{Jones-Lederman-OUP-2006, DarianSmith-HandbookPhysiology-1984, Johansson-Flanagan-NatRevNeurosci-2009}. In doing so, it filters and shapes the tactile information thus effectively participating in the signal processing and tactile encoding processes. In 1981, Phillips and Johnson developed a semi-infinite elastic model of the digital skin to predict the stress and strain state at the location of mechanoreceptors terminations produced by indenting the skin surface with a substrate of known topography \cite{Phillips-Johnson-JNeurophysiol-1981b, Phillips-Johnson-JNeurophysiol-1981c}. This simple biomechanical model allowed them to show that certain characteristics of slowly adapting (SA1) mechanoreceptors response (such as edge enhancements) could be ascribed to the mechanical (pre-neural) stage of the tactile transduction process. 
 
In this early model, the skin surface was considered smooth. This approach thus ignored the epidermal ridges (fingerprints) which characterize the digital skin of primates. These surface microstructures have long been suspected to favor tactile perception, although the mechanism at play remained poorly understood \cite{Loesch-Martin-AnnHumBiol-1984}. In recent years, the development of numerical and artificial models of the fingertip have allowed for the understanding of how these wavy structures modify the stress distribution within the skin \cite{Gerling-Thomas-EC-2005, Maeno-Kobayashi-Yamazaki-JSMEIntJ-1998, Yamada-Maeno-Yamada-JRobMechatron-2002, Vasarhelyia-Fodorb-Roska-SensorsActuatorsA-2007}. These studies gave a solid basis to an earlier proposition by Cauna that each epidermal ridge, as a single unit, might act as an arm-lever and increase the subcutaneous strain \cite{Cauna-AnatRec-1954}. In order for this stress focusing process to be functional, however, the mechanoreceptor termination needs to lie relatively close to the skin surface and at precise locations with respect to the epidermal ridge \cite{Maeno-Kobayashi-Yamazaki-JSMEIntJ-1998}. This is the case for SA1 receptors \cite{Vallbo-Johansson-Pergamon-1978}, which mediate the perception of coarse textures, but not for the deeper Pacinian corpuscles which are implicated in the coding of finer textures (lateral size below $\sim$200$\mu$m) \cite{Hollins-Bensmaia-CanJExpPsy-2007}.

In a recent article, we have suggested however that fingerprints may also contribute to the perception of fine textures through a completely different mechanism \cite{Scheibert-Leurent-Prevost-Debregeas-Science-2009}. This perception is mediated by the rapid skin vibrations elicited when actively scanning the fingertip over the surface. We have shown that the presence of epidermal ridges spatially modulates the interfacial stress field between the skin and the substrate. This in turn results in an amplification of the subcutaneous skin vibrations induced by textural components of period similar to that of the ridges themselves. This process requires the mechanoreceptor's receptive field to be much larger than the inter-ridge period as is the case for PCs.  It is weakly dependent on the mechanoreceptor location with respect to the fingerprints pattern. Also, the ridges exact profile should not matter provided that (i) they are deep enough to induce large interfacial stress modulations, (ii) they are oriented perpendicular to the scanning axis.

This mechanism was demonstrated using a biomimetic sensor that mimics the operation of a single PC in the fingertip. One legitimate question is whether the mechanism evidenced with this idealized device is relevant to an actual fingertip. Answering this question poses a technical challenge since there is no current way to measure the stress experienced by mechanoreceptors in an actual finger. However, one may expect that differences in texture-induced subcutaneous vibrations may show up in the global friction force acting on the finger during tactile exploration. This idea is at the basis of recent experiments in which the vibrations of the fingertip skin were locally measured with a displacement probe as the finger was scanned across various textured substrates \cite{Hollins-Bensmaia-CanJExpPsy-2007}. The intensity of these vibrations weighted by the PCs' spectral sensitivity could be correlated with the perception of roughness determined independently through psychophysical experiments \cite{Bensmaia-Hollins-PerceptionPsychophysics-2005}.

\begin{figure}
\includegraphics[width=\columnwidth]{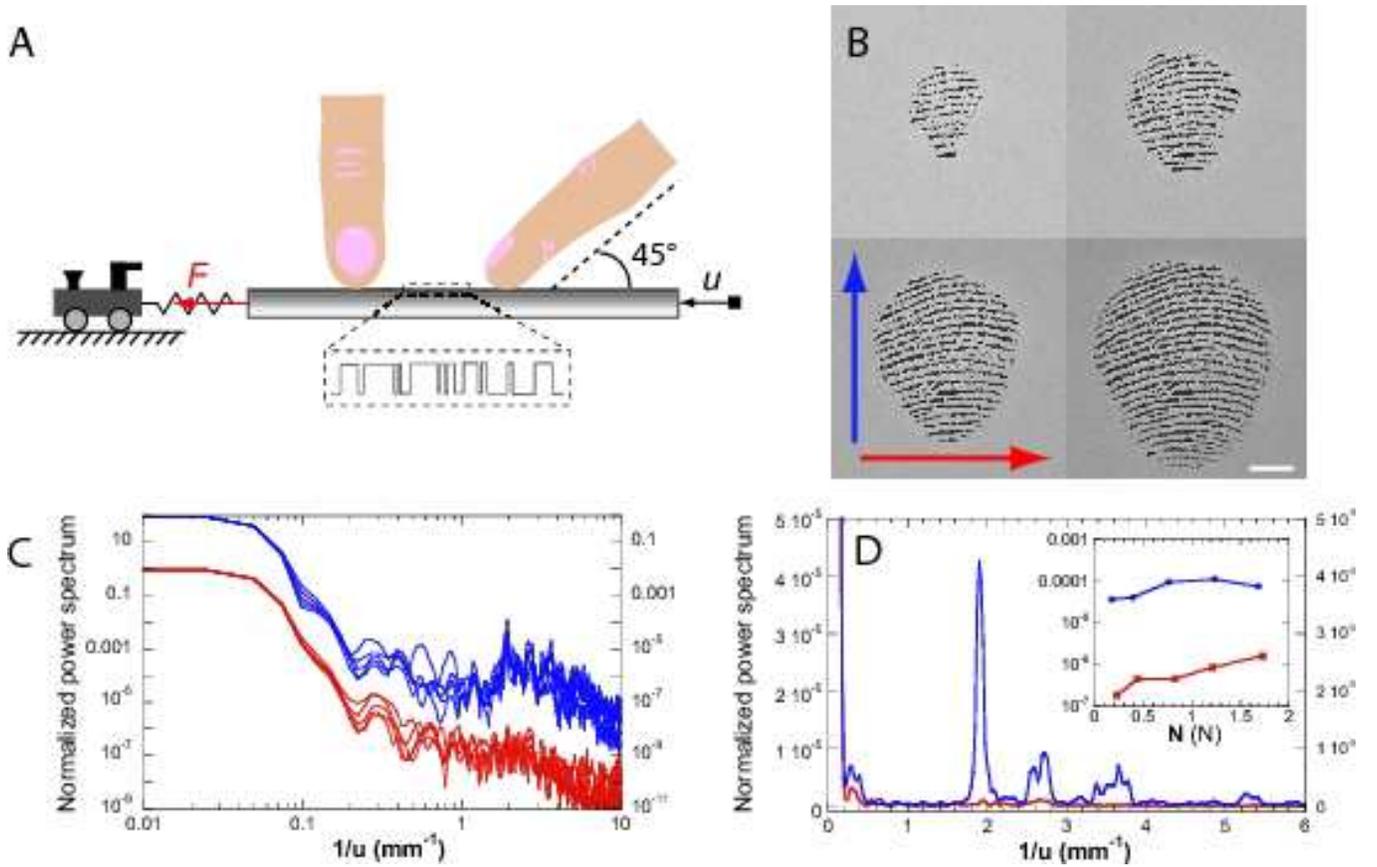}
\caption{\textbf{A} Experimental set-up. The substrate (in grey) consists of a 50mm long glass slide patterned with 1D rough textures (see ref. \cite{Scheibert-Leurent-Prevost-Debregeas-Science-2009} for details). It is mounted on a double cantilever apparatus which allows one to monitor both the normal and tangential forces (denoted $N$ and $F$, respectively) acting on it. The forefinger (male, 37 years old) is constrained in a fixed position at 45$^\circ$ with respect to the surface. Once in place, the substrate is brought in contact at a prescribed load $N$. It is then moved at a constant speed $v$=5mm/s using a DC-current motor along either distal (mode 1, right) or radial (mode 2, left) direction. \textbf{B} Typical images of the contact zone between the fingertip and a smooth substrate for 4 different normal loads ($N$=0.2, 0.4, 0.8, 1.6N). The finger is pointing upward. The arrows correspond to the scanning direction in mode 1 (blue) and mode 2 (red). The white bar is 2mm long. \textbf{C} Comparison of the normalized power spectra of the tangential force $F$ obtained in mode 1 (blue, shifted vertically for clarity) and mode 2 (red). The different graphs correspond to loads $N$=0.2, 0.4, 0.8, 1.2, 1.7N. \textbf{D} Linear/linear plot of the spectra for $N$=0.4N. The maximum of the normalized power spectrum in mode 1 occurs at the spatial frequency 2mm$^{-1}$. Inset: spectrum amplitude at this particular spatial frequency as a function of $N$. \label{FigCIB}}
\end{figure}

Along a similar line, we have designed a set-up, sketched in Figure 1A, which allows one to monitor the tangential force acting on a flat substrate as it is rubbed against a fingertip at constant speed and normal load. The stimuli substrates, described in ref. \cite{Scheibert-Leurent-Prevost-Debregeas-Science-2009}, consist of 50mm long glass slides patterned with square wave gratings whose edges positions are randomly distributed, thus resulting in a white noise texture. The forefinger is constrained in a fixed position at 45$^\circ$ with respect to the substrate plane. In this configuration, the fingerprints in the contact zone have a preferred orientation as shown in Figure 1B. By comparing the force signal obtained upon moving the substrate distally (mode 1) versus radially (mode 2), one can thus directly probe the impact of fingerprints in shaping the skin vibrations spectrum. 

Figure 1C displays the power spectra of vibrations obtained in mode 1 and mode 2 for different normal loads $N$ in the range 0.2N<$N$<1.7N and scanning speed $v$=5mm/s, as a function of the inverse of the substrate displacement $u$=$vt$. Both scanning orientations produce comparable spectra at low frequency but the mode 1 yields larger amplitude of vibrations for spatial frequencies in the range 1.5-5mm$^{-1}$. As shown in Figure 1D, the maximum amplification corresponds to a spatial frequency equal to the inverse of the inter-ridge distance $\lambda$ which is of the order of 0.5mm. The inset of figure 1D shows the dependence of the amplitude of the power spectrum at 1/$\lambda$ with $N$. Over the range of loads explored, the relative amplification induced by the fingerprints is of the order of 100.  

This simple biomechanical experiment confirms the relevance for human tactile perception of the mechanism of spectral amplification evidenced in ref. \cite{Scheibert-Leurent-Prevost-Debregeas-Science-2009} using a biomimetic approach. It shows that the relative orientation of the scanning axis and the fingerprints in contact with the surface determines to a large extent the spectrum of skin vibrations elicited during tactile exploration. This work indicates that comparing radial and distal scanning orientations provides a simple but efficient way to test the role of fingerprints in shaping texture induced skin vibrations. The same approach could be similarly implemented in neuro- or psychophysical experiments.


\end{document}